\documentclass[showpacs,twocolumn,pre]{revtex4}
\usepackage{amsmath}

\input{tcilatex}

\begin{document}

\title{Bethe ansatz solution of zero-range process with nonuniform
stationary state}
\author{A. M. Povolotsky$^*$}

\begin{abstract}
The eigenfunctions and eigenvalues of the master-equation for zero range
process with totally asymmetric dynamics on a ring are found exactly using
the Bethe ansatz weighted with the stationary weights of particle
configurations. The Bethe ansatz applicability requires the rates of hopping
of particles out of a site to be the $q$-numbers $[n]_q$. This is a
generalization of the rates of hopping of noninteracting particles equal to
the occupation number $n$ of a site of departure. The noninteracting case
can be restored in the limit $q\to 1$. The limiting cases of the model for $%
q=0,\infty$ correspond to the totally asymmetric exclusion process, and the
drop-push model respectively. We analyze the partition function of the model
and apply the Bethe ansatz to evaluate the generating function of the total
distance travelled by particles at large time in the scaling limit. In case
of non-zero interaction, $q \ne 1$, the generating function has the
universal scaling form specific for the Kardar-Parizi-Zhang universality
class.
\end{abstract}

\pacs{05.40.-a, 02.50.-r, 64.60.Ht}
\affiliation{Physics Department, University of Aveiro, Campus de Santiago - 3810-193,
Aveiro, Portugal}
\affiliation{Bogoliubov Laboratory of Theoretical Physics, Joint Institute for Nuclear
Research, Dubna 141980, Russia}
\email{povam@thsun1.jinr.ru}
\date{\today }
\maketitle

\section{ Introduction}

The Bethe ansatz \cite{bethe} is one of the most powerful tools to get exact
results for the systems with many interacting degrees of freedom in low
dimensions. The exact solutions of one-dimensional quantum spin chains and
two-dimensional vertex models are classical examples of its application \cite%
{baxter}. In the last decade, the Bethe ansatz was shown to be useful to
study one-dimensional stochastic processes \cite{gs,schutz}. The first and
most explored example is the asymmetric simple exclusion process (ASEP),
which serves as a testing ground for many concepts of the nonequilibrium
statistical physics \cite{derrida}. Yet several other Bethe ansatz solvable
models of nonequilibrium processes have been proposed such as the 
asymmetric diffusion models \cite{sw,sw2}, generalizations of the drop-push
model \cite{srb,karim1,karim2}, and the asymmetric avalanche process (ASAP) 
\cite{piph}.

Most of models studied by the coordinate Bethe ansatz have a common
property. That is, a system evolves to the stationary state, where all the
particle configurations occur with the same probability. This property can
be easily understood from the structure of the Bethe eigenfunction. Indeed,
the stationary state is given by the groundstate of evolution operator,
which is the eigenfunction with zero eigenvalue and momentum. Such Bethe
function does not depend on particle configuration at all and results in the
equiprobable ensemble. Except for a few successful attempts to apply the
Bethe ansatz to systems with nonuniform stationary state, such as the ASEP
with blockage \cite{schutz1} or defect particle \cite{de}, there is no  much
progress in this direction.

On the other hand many interesting physical phenomena such as condensation
in nonequilibrium systems \cite{evans}, boundary induced phase transitions 
\cite{Krug,sd} or intermittent-continuous flow transition \cite{pph3} become
apparent from non-trivial form of the stationary state, which change
dramatically from one point of phase space to another. Typical example is
the zero-range process (ZRP), served as a prototype of a one-dimensional
nonequilibrium system exhibiting the condensation transition \cite{evans}.
While its stationary measure has been studied in detail \cite{gss}, the full
dynamical description is still absent.

The aim of this paper is to obtain the Bethe ansatz solution of the ZRP. The
paper is organized as follows. In section II we use the Bethe ansatz to
solve the eigenfunction and eigenvalue problem for the master equation of
the ZRP and show that its applicability requires the rates of hopping of
particles out of a site to be the $q$-numbers. We show that with this choice
of the rates the model is equivalent to the q-boson totally asymmetric
diffusion model. In section III we obtain the partition function of the ZRP
with the rates obtained and evaluate some stationary correlations. In
section IV we apply the equations obtained from the Bethe ansatz solution to
derive the expression for the generating function of the total distance
travelled by particles in the large time limit. We summarize the results in
section V.

\section{Master equation of zero range process}

Let us consider the system of $p$ particles on a ring consisting of $N$
sites. Every site can hold an integer number of particles. Every moment of
time, one particle can leave any occupied site, hopping to the next site
clockwise. The rate of hopping $u(n_{i})$ depends only on the occupation
number $n_{i}$ of the site of departure $i$. The stationary measure of such
a process has been found to be a product measure \cite{evans}, i.e. the
probability of configuration $\{n_{i}\}$, specified by the occupation
numbers $\{n_{1},n_{2},\ldots ,n_{N}\}$, is, up to the normalization factor,
given by the weight 
\begin{equation}
W\left( \{n_{i}\}\right) =\prod_{i=1}^{N}f(n_{i}),  \label{weights}
\end{equation}%
where one-site weight $f(n)$ is 
\begin{equation}
f(n)=\prod_{m=1}^{n}\frac{1}{u(m)}  \label{f(n)}
\end{equation}%
if $n>0$, and $f(0)=1$.

Let us consider the probability $P_{t}(n_{1},\ldots ,n_{N})$ for $N$ sites
to have occupation numbers $n_{1},\ldots ,n_{N}$ at time $t$. It obeys the
master equation defined by the dynamics described. 
\begin{widetext}
\begin{equation}\label{master}
 \partial_t P_{t}(n_{1},\ldots ,n_{N})= \sum_{\begin{matrix}
    k=1\\
     n_k\ne0
  \end{matrix}}^{N}\left(u(n_{k-1}+1)P_{t}(\ldots,n_{k-1}+1,n_k-1,\ldots)-
u(n_{k})P_{t}(n_{1},\ldots,n_{N})\right)
\end{equation}
\end{widetext}
Here, cyclic boundary condition, $n_{-1}\equiv n_{N}$, is imposed.

Another way to specify the configuration of system is to use the set of
coordinates of $p$ particles $\{x_{i}\}=\{x_{1},\ldots ,x_{p}\}$, rather
than occupation numbers $\{n_1,\dots,n_N\}$, the two ways being completely
equivalent. While the explicit form of the master equation in such notations
is more complicated, it turns out more appropriate for an analytic solution.
The main idea of the solution is to use the Bethe ansatz for the function $%
P^0_{t}(x_{1},\ldots ,x_{p})$ related with the solution of the master
equation $P_{t}(x_{1},\ldots ,x_{p})$ as follows 
\begin{equation}
P_{t}(x_{1},\ldots ,x_{p})=W(\{n_{i}\})P_{t}^{0}(x_{1},\ldots ,x_{p}).
\label{solution}
\end{equation}

\subsection{Two-particle sector}

To explain the details we first consider the case $p=2$ subsequently
generalizing it to the case of arbitrary $p$. Without lost of generality we
can define 
\begin{equation}  \label{u(1),u(2)}
u(1)\equiv 1,\,\,\,\,u(2)\equiv u>0.
\end{equation}

Now, we are going to show that the solution, $P_{t}^{0}(x_{1},x_{2})$, of
the master equation for noninteracting particles, 
\begin{eqnarray}
\partial_{t}P_{t}^{0}(x_{1},x_{2})=
P_{t}^{0}(x_{1}-1,x_{2})&+&P_{t}^{0}(x_{1},x_{2}-1)  \notag \\
&-&2P_{t}^{0}(x_{1},x_{2}),  \label{master02}
\end{eqnarray}%
is related with the solution $P_{t}(x_{1},x_{2})$ of the master equation for
the ZRP in the domain $x_1\le x_2$ through the relation (\ref{solution}),
provided that the former satisfy certain constraint. Indeed, when $\left(
x_{2}-x_{1}\right) \geq 2$, the equation for probability $P_{t}(x_{1},x_{2})$
for the ZRP coincides with one for noninteracting particles (\ref{master02}%
). For $x_{1}=x_{2}-1=x$ the equation for the ZRP%
\begin{equation}
\partial _{t}P_{t}(x,x+1)=P_{t}(x-1,x)+uP_{t}(x,x)-2P_{t}(x,x+1)
\label{P(x,x+1)eq}
\end{equation}%
can be obtained from (\ref{master02}) if we define 
\begin{equation}
P_{t}(x,x)=\frac{1}{u}P_{t}^{0}(x,x),  \label{P(x,x)}
\end{equation}%
which is consistent with (\ref{solution}). In the case $x_{2}=x_{1}=x$ the
equation (\ref{master02}) contains the term $P_{t}^{0}(x,x-1)$ which is
beyond the "allowed" region $x_1\le x_2$ and, thus, does not carry any
physical content. To restore the correct equation for the ZRP 
\begin{equation}
\partial _{t}P_{t}(x,x)=P_{t}(x-1,x)-uP_{t}(x,x),  \label{P(x,x)eq}
\end{equation}
we can redefine this term to compensate the difference between the equations
(\ref{master02}) and (\ref{P(x,x)eq}). As a result we get the following
constraint on $P_{t}^{0}(x_1,x_2)$ 
\begin{equation}
P_{t}^{0}(x,x-1)=(u-1) P_{t}^{0}(x-1,x)-\left( u-2\right) P_{t}^{0}(x,x).
\label{constraint}
\end{equation}
Thus, the solution of the free equation (\ref{master02}), which satisfies
the constraint (\ref{constraint}), gives the solution of the master equation
for the ZRP in the domain $x_1\le x_2$. Now, we can use the Bethe ansatz for
the eigenfunction of the free equation (\ref{master02}) 
\begin{equation}  \label{p=2,Bethe ansatz}
P_{t}^{0}(x_{1},x_{2})=e^{\lambda t}\left(
A_{1,2}z_{1}^{-x_{1}}z_{2}^{-x_{2}}+A_{2,1}z_{1}^{-x_{2}}z_{2}^{-x_{1}}%
\right).
\end{equation}
Its substitution to the equation (\ref{master02}) results in the expression
for the eigenvalue. 
\begin{equation}
\lambda =z_{1}+z_{2}-2.
\end{equation}
The ansatz (\ref{p=2,Bethe ansatz}) to be consistent with the constraint (%
\ref{constraint}), the amplitudes $A_{1,2}$ and $A_{2,1}$ should satisfy the
relation 
\begin{equation}
\frac{A_{1,2}}{A_{2,1}}=-\frac{(2-u)-(1-u)z_2-z_1}{(2-u)-(1-u)z_1-z_2},
\end{equation}
which together with the cyclic boundary conditions, $%
P_t^0(x_1,x_2)=P_t^0(x_2,x_1+N)$, results in the system of two algebraic
equations. The first one is the following 
\begin{equation}
z_1^{-N}=-\frac{(2-u)-(1-u)z_2-z_1}{(2-u)-(1-u)z_1-z_2},
\end{equation}
while the second can be obtained by the change $z_1\leftrightarrow z_2$.

\subsection{Many-particle sector}

To generalize these results to the case of arbitrary $p$ let us consider the
configuration with two neighboring sites $(x-1)$ and $x$ having occupation
numbers $m$ and $n$ respectively. Let us explicitly write down the terms of
the master equation corresponding to the transition into and from this
configuration due to a particle jump into and from the site $x$
respectively, 
\begin{widetext}
\begin{eqnarray}
\partial _{t}P_{t}(\ldots ,(x-1)^{m},(x)^{n},\ldots )=
\ldots +u(m+1)P_{t}(\ldots ,(x-1)^{m+1},(x)^{n-1},\ldots )\notag \\ -u(n)P_{t}(\ldots ,(x-1)^{m},(x)^{n},\ldots
)+\ldots . \label{masterZRP}
\end{eqnarray}%
Here $(x)^{n}$ denotes $n$ successive arguments $x$ of the function $P_{t}(x_{1},\ldots
,x_{p})$. In terms of $P_{t}^{0}(x_{1},\ldots,x_{p})$,
which is related to $P_{t}(x_{1},\ldots ,x_{p})$ according to (\ref%
{solution}), this equation looks as follows
\begin{eqnarray}
\partial _{t}P_{t}^{0}(\ldots ,(x-1)^{m},(x)^{n},\ldots )=
\ldots +u(n) \times\left[ P_{t}^{0}(\ldots ,(x-1)^{m+1},(x)^{n-1},\ldots )\right.\notag\\- \left. P_{t}^{0}(\ldots
,(x-1)^{m},(x)^{n},\ldots )\right] +\ldots \hspace{-3mm}. \label{masterZRP_P^0}
\end{eqnarray}%
Note, that in this form the coefficient before the term in square brackets  is equal
to $u(n)$, i.e. does not depend on the number $m$ of particles in the site $%
(x-1)$. Thus, the r.h.s of the master equation expressed through $%
P_{t}^{0}(x_{1},\ldots ,x_{p})$ is the sum of one-site factors similar to those in
(\ref{masterZRP_P^0}) for all nonempty sites. Equating such term with one from the equation
for noninteracting  particles,
\begin{equation}
 \partial _{t}P_{t}^{0}(x_{1},\ldots ,x_{p})=  \label{free-p}  \sum_{i=1}^{p}\left( P_{t}^{0}(\ldots
,x_{i}-1,\ldots )-P_{t}^{0}(\ldots ,x_{i},\ldots )\right)
\end{equation}%
we obtain the following constraint for $P_{t}^{0}(x_{1},\ldots ,x_{p})$,
\begin{eqnarray}
(u(n)-1)P_{t}^{0}(\ldots ,(x),(x+1)^{n-1},\ldots )-  \sum_{j=2}^{n}P_{t}^{0}(\ldots
,(x+1)^{j-1},(x),(x+1)^{n-j},\ldots )\notag\\-(u(n)-n)P_{t}^{0}(\ldots ,(x+1)^{n},\ldots )=0. \label{p-constraint}
\end{eqnarray}%
\end{widetext}
Such relations for all $n$ are to be understood as a redefinition for the
terms outside of allowed region $x_{1}\leq \ldots \leq x_{p}$. The Bethe
ansatz to be applicable, the relation (\ref{p-constraint}) should be
reducible to the two particle constraint (\ref{constraint}). This can be
proved by induction. To this end, we assume that similar relations including 
$u(k)$ are reducible for all $k<n$. Then starting from the relation (\ref%
{p-constraint}), which includes rate $u(n)$, we apply (\ref{constraint}) to
every pair $(x,x-1)$ of arguments of the function $P_{t}^{0}(\ldots )$ under
the sum and require the result to be similar relation for $u(n-1)$. We
obtain the following recurrent formula for the rates 
\begin{equation}
u(n)=1-(1-u)u(n-1),  \label{u(n)}
\end{equation}%
which can be solved in terms of $q$-numbers 
\begin{equation}
u(n)=[n]_{q}=\frac{1-q^{n}}{1-q}\,\,\,,  \label{u(n)=[n]}
\end{equation}%
where 
\begin{equation}
q=u-1.  \label{[n_q]}
\end{equation}%
Further generalization of two-particle results is straightforward. We use
the Bethe ansatz for the eigenfunction $P_{t}^{0}(x_{1},\ldots ,x_{p})$ of
the equation (\ref{free-p}) 
\begin{equation}
P_{t}^{0}(x_{1},\ldots ,x_{p})=e^{\lambda t}\sum_{\{\sigma _{1},\ldots
,\sigma _{p}\}}A_{\{\sigma _{1},\ldots ,\sigma
_{p}\}}\prod_{i=1}^{p}z_{\sigma _{i}}^{-x_{i}}  \label{bethe ansatz}
\end{equation}%
Here $z_1,\ldots, z_p$ are complex numbers, the summation is taken over all $%
p!$ permutations $\{\sigma _{1},\dots ,\sigma _{n}\}$ of the numbers $%
(1,\ldots ,p)$, and the coordinates of particles are ordered in the
increasing order $x_{1}\leq x_{2}\leq \ldots \leq x_{p}$.

Substituting (\ref{bethe ansatz}) into (\ref{free-p}) we obtain the
expression for the eigenvalue, 
\begin{equation}
\lambda =\sum_{i=1}^{p}z_{i}-p.  \label{eigenvalue}
\end{equation}%
The numbers $(z_{1},\ldots ,z_{p})$ are to be defined from the Bethe
equations, 
\begin{equation}
z_{i}^{-N}=(-1)^{p-1}\prod_{j=1}^{p}\frac{(2-u)-(1-u)z_{j}-z_{i}}{%
(2-u)-(1-u)z_{i}-z_{j}},  \label{Bethe equations}
\end{equation}%
which follow from the condition of compatibility of cyclic boundary
conditions with the constraint (\ref{constraint}).

For the sake of convenience in the following discussion we use the parameter 
$q$ defined in (\ref{[n_q]}) rather than $u$. We should note that the
appearance of q-numbers as the condition of the Bethe ansatz integrability
is not unexpected. The notion of q-deformation naturally appears in context
of Bethe ansatz solvable models characterized by the trigonometric R-matrix.
In algebraic language this is the consequence of the of the fundamental
Yang-Baxter equation which leads to q-commutation relations between the
local operators constituting the transfer matrices \cite{faddeev}. One of
such models, q-boson totally asymmetric diffusion model \cite{sw2}, turns
out to be directly related to the model we consider. In order to see the
correspondence, let us formally write the distribution $P_t(C)$  as a vector
of state 
\begin{equation}
\left|P_t(C)\right\rangle=\sum_{\{C\}}P_t(C)\left|C\right\rangle,
\end{equation}
where $C$ is a configuration of particles on the lattice, $%
C=\{n_1,\ldots,n_N \}$, and the summation is over all configurations.
Consider the algebra generated by the operators ${B_j,B_j^+},\mathcal{N}_j$,
which act on the occupation number $n_j>0$ of each site $j$ of the lattice
as follows: 
\begin{eqnarray}  \label{B^+}
B_j \left |n_j\right\rangle &=& \left |n_j-1\right\rangle \\
B^+_j \left |n_j\right\rangle &=& [n_j+1]_q\left |n_j+1\right\rangle \\
\mathcal{N}_j \left |n_j\right\rangle &=& n_j \left|n_j\right\rangle,
\end{eqnarray}
The state $\left |0\right\rangle$ plays the role of vacuum state 
\begin{equation}
B_j\left |0_j\right\rangle=0.
\end{equation}
Then, the master equation (\ref{master}) with the rates given by (\ref%
{u(n)=[n]}) can be written in the from of the imaginary time Schr$\ddot{%
\mathrm{o}}$dinger equation 
\begin{equation}
\partial_t \left|P_t(C)\right\rangle=-\mathbf{H}\left|P_t(C)\right\rangle
\end{equation}
where the hamiltonian \textbf{H} is given in terms of the operators (\ref{B},%
{\ref{B^+}) 
\begin{equation}
\mathbf{H}=-\sum_j (B^+_{j-1}B_j-B^+_jB_j).  \label{hamiltonian}
\end{equation}
One can directly check that the operators ${B_j,B_j^+},\mathcal{N}_j$
satisfy the following commutation relations 
\begin{eqnarray}  \label{commutation1}
\left[B_j,B^+_k \right]&=&q^{\mathcal{N}_j} \delta_{jk}, \\
\left[\mathcal{N}_k,B_j\right]&=&-B_j \delta_{jk} \\
\left[\mathcal{N}_k,B^+_j\right]&=&B^+_j \delta_{jk} .  \label{commutation3}
\end{eqnarray}
These commutation relations and  the hamiltonian (\ref{hamiltonian}) give
us, up to the change of notations $q \to q^{-2}$, the definition of the
q-boson totally asymmetric diffusion model. Obviously, the dynamical rules
of the ZRP with the rates given by (\ref{u(n)=[n]}) is nothing but the
explicit realization of this hamiltonian. Its integrability has been shown
in \cite{sw2} via algebraic Bethe ansatz and two particle diffusion on the
infinite lattice has been considered. Note that while in \cite{sw2} the
q-boson totally asymmetric diffusion model was initially defined in terms of
q-boson operators, we started from the ZRP with arbitrary rates and come to
q-numbers as the integrability condition. }

Let us first take a qualitative look at the behaviour of the ZRP resulted by
the choice (\ref{u(n)=[n]}) of the rates $u(n)$ for different values of $q$.
In the limit $q \to 1$ the $q$-numbers degenerate into simple numbers.
Therefore, the rates are given by $u(n) = n$, which corresponds to the
diffusion of noninteracting particles. The Bethe equations in this case
decouple to the form $z_j^N=1$ as is expected in noninteracting case. In the
domain $q >1$ the rates $u(n)$ grow exponentially with $n$, which
corresponds to the interaction between particles effectively accelerating
free diffusive motion, i.e. the higher density of particles the faster is
their mean velocity. In the limit $q\to \infty $ the model is equivalent to $%
n = 1$ drop push model \cite{srb}, which is confirmed by the same form of
Bethe equations \cite{karim1}. In the domain $0<q < 1$ the rates $u(n)$ grow
monotonously from $(q+1)$ for $n=2$ to $1/(1-q)$ for $n\to\infty$, resulting
in the interaction between particles slowing down the particle flow
comparing to the one of noninteracting diffusing particles. When $q=0$, all
the rates do not depend on the number of particles, i.e. $u(n) = 1$. This
case, (also referred to as phase model \cite{BIK}), can be mapped on the
totally asymmetric ASEP by insertion of one extra bond before every
particle. At last, in the domain $-1<q<0$ the rates $u(n)$ also saturate to
the constant $1/(1-q)$ with growth of $n$, though oscillating around this
value. As it has been mentioned above, the ZRP served as an example of the
nonequilibrium system with the condensation transition. In our case however
the condensation is absent as the rates $u(n)$ defined above do not satisfy
Evans criteria, according to which the condensation in the ZRP occurs if the
rates saturate to a constant slower than $2/n$.

It is interesting, that the recursion relation (\ref{u(n)}) rewritten in
terms of the quantity $(1-u(n))$ coincide with one for the toppling
probabilities $\mu_n$, imposed by the requirement of the Bethe ansatz
integrability in the ASAP. In fact, the ASAP can be represented as a special
case of the discrete time ZRP viewed from the reference frame moving
together with an avalanche. This situation, however, is quite different from
one considered here. In the moving reference frame all particles hop
definitely to the next site except of one from an active site, which is the
only site with multiple occupation. In that case quantity $1-\mu_n$ plays
the role of probability of hopping of a particle out of this site. This
dynamics leads to the picture  similar to the ZRP on an inhomogeneous
lattice with one attractive site. Such ZRP was shown to exhibit the
condensation transition. In terms of the avalanche processes it is the
transition from the intermittent to continuous flow. Despite the nonuniform
stationary state of the ASAP in discrete time \cite{pph3}, its Bethe ansatz
solution was based on the continuous time picture considered on the ensemble
of equiprobable configurations with at most one particle occupation. The
avalanches were accounted in the rates of transitions between these
configurations, generating infinite series in the master equation.

To use the eigenfunctions obtained for the construction of particular
solutions one should first question if they form complete orthogonal basis.
In general this question is not easy to answer, as the set of solutions of
the Bethe equations is not known. Some arguments have been given, \cite%
{gs,kim}, for the the totally asymmetric exclusion process due to its
connection with the asymmetric six-vertex model\cite{baxter1}. The long time
characteristics of the process can, however, be analysed without discussing
this question. To this end, we can use the properties of the largest
eigenvalue for slightly modified equation, which describes the generating
function of total distance travelled by particles, \cite{dl}. The advantages
of this approach are first, that the uniqueness of the largest eigenvalue is
guarantied by Perron-Frobenius theorem. Second, corresponding solution of
the Bethe equations can be easily identified as it corresponds to the
stationary state of the process. To give an example of application of the
above results we perform this analysis in the section IV.

\section{Structure of the stationary state}

Before going to the results of the analysis of the Bethe equations, let us
first look at the structure of the stationary measure of the model. It is
characterized by the partition function 
\begin{eqnarray}
Z(N,p)=\hspace{-5mm}\sum_{\{n_{1},\ldots ,n_{p}=1\}}^{\infty }\hspace{-5mm}%
\delta (n_{1},+\ldots +n_{N}-p)\prod_{i=1}^{N}{f(n_{i})} ,  \label{Z(N,p)}
\end{eqnarray}%
where one site weight $f(n)=1/[n]_{q}!$, defined in (\ref{f(n)}), is
expressed through the $q$-factorial $[n]_{q}!=\prod_{k=1}^{n}[k]_{q}$. In
the limit $q\rightarrow 1$, $q$-factorial turns into simple factorial, as it
should be in the noninteracting case. The sum in (\ref{Z(N,p)}) can be
presented in form of the contour integral 
\begin{equation}
Z(N,p)=\frac{1}{2\pi i}\oint \frac{\left( F(z)\right) ^{N}}{z^{p+1}}dz,
\label{Z(N,p),int}
\end{equation}%
where the series $F(z)=\sum_{n=0}^{\infty }f(n)z^{n}$ can be summed to the
infinite product due to the $q$-binomial theorem \cite{aar}: for $|q|<1$ 
\begin{equation}
F(z)=\sum_{n=0}^{\infty }\frac{\left( z(1-q)\right) ^{n}}{(q;q)_{n}}%
=(z(1-q);q)_{\infty }^{-1}  \label{F(z),|q|<0}
\end{equation}%
and for $|q|>1$ 
\begin{eqnarray}
F(z) &=&\sum_{n=0}^{\infty }\frac{\left( z(1-q^{-1})\right) ^{n}q^{-n\left(
n-1\right) /2}}{(q^{-1};q^{-1})_{n}}  \label{F(z),|q|>0} \\
&=&(z(q^{-1}-1);q^{-1})_{\infty }.  \notag
\end{eqnarray}%
Here, we used the notation $(a;q)_{n}=\prod_{k=0}^{n-1}(1-aq^{k})$ for
shifted $q$-factorial. The above $q$-series are known as $q$-analogs of the
exponential function $e^{z}$, which can be restored in the limit $q\to 1$.
The presence of $q$-analogs of the functions, which appear in the case of
noninteracting particles, is a direct consequence of the replacement of
simple numbers by $q$-numbers in the expression of the rates of hopping. In
the thermodynamic limit, $N\rightarrow \infty ,p\rightarrow \infty ,p/N=\rho 
$, the integral (\ref{Z(N,p),int}) can be calculated in the saddle point
approximation. The equation for the saddle point $z_{0}$, 
\begin{equation}
\rho =z_0\log ^{\prime }F(z_{0}),
\end{equation}%
contains the logarithmic derivative of $F(z)$, which can be evaluated using
the product form of $F(z)$ (\ref{F(z),|q|<0},\ref{F(z),|q|>0}). As a result
we obtain the following relations between $z_0,\rho$ and $q$. 
\begin{equation}
\rho =z_{0}(1-q)\sum_{n=0}^{\infty }\frac{q^{n}}{1-z_{0}(1-q)q^{n}}
\label{saddle point,
|q|<1}
\end{equation}%
\begin{equation}
\rho =z_{0}(1-q^{-1})\sum_{n=0}^{\infty }\frac{q^{-n}}{%
1-z_{0}(q^{-1}-1)q^{-n}}  \label{saddle point, |q|>1}
\end{equation}%
for $|q|<1$ and $|q|>1$ respectively. Below the same equations will appear
in a different context from the analysis of Bethe ansatz equations. Then,
the partition function, 
\begin{equation}
Z(N,p)=\frac{\left(F(z_{0})\right)^N}{z_{0}^{p }},
\end{equation}
can be used to calculate stationary state correlations such as the speed, $%
v=\langle u(n)\rangle $, i.e. the average hopping rate out of a site 
\begin{equation}
v=\frac{Z(N,p-1)}{Z(N,p)}=z_{0}  \label{v}
\end{equation}%
or the probability distribution of the number of particles in a site 
\begin{equation}
P(n)=f(n)\frac{Z(N-1,p-n)}{Z(N,p)}=\frac{1}{[n]_{q}!}\frac{z_{0}^{n}}{%
F(z_{0})}.  \label{P(n)}
\end{equation}

\section{ The long time behaviour from the Bethe equations}

To obtain any results beyond the stationary correlations one needs to
analyze the above Bethe ansatz solution. Since similar analysis has been
done several times before \cite{bs,kim,lk,pph2}, we do not give the detailed
calculations here. Instead we outline the main points of the solution to
emphasize the connection with the formulas obtained from the analysis of
stationary measure.

Consider the generating function $\mathcal{F}_{t}(C)=\sum_{Y=0}^{\infty
}P_{t}(C,Y)e^{\gamma Y}$, where $P_{t}(C,Y)$ is the joint probability for
the system to be in the configuration $C$ at time $t$ the total distance
travelled by particles being $Y$. The only difference of the equation for $%
\mathcal{F}_{t}(C)$ from the equation (\ref{master}) is the coefficient $%
e^{\gamma }$ before the first term under the sum in the r.h.s., which
corresponds to the increasing of travelled distance by one due to the
hopping of one particle. At large time, $t\rightarrow \infty $, the
behaviour of the generating function of the distance $Y_{t}$ travelled by
particles up to time $t$, $\langle e^{\gamma Y_{t}}\rangle =\sum_{C}\mathcal{%
F}_t(C)$, is determined by the largest eigenvalue of the equation for $%
\mathcal{F}_t(C)$. 
\begin{equation}
\lambda _{0}(\gamma )=\lim_{t\to \infty }\frac{\log \langle e^{\gamma
Y_{t}}\rangle}{t}
\end{equation}
Using the ansatz (\ref{solution},\ref{bethe ansatz}) for the eigenfunction
of the equation for $\mathcal{F}_t(C)$ we can repeat all the above
arguments. Then, if we make the variable change $x_{i}=1-e^{\gamma }z_{i}$,
the eigenvalue and the Bethe equations will simplify to the following form. 
\begin{equation}
\lambda (\gamma )=-\sum_{i=1}^{p}x_{i}  \label{eigenvalue_gamma}
\end{equation}%
\begin{equation}
e^{\gamma N}(1-x_{i})^{-N}=(-1)^{p-1}\prod_{j=1}^{p}\frac{x_{i}-qx_{j}}{%
x_{j}-qx_{i}}  \label{Bethe equations_gamma}
\end{equation}%
In these variables the r.h.s. of the Bethe equations coincides with one for
the partially asymmetric ASEP and the ASAP. This allows us to modify the
techniques developed for the analysis of these processes.

In the thermodynamic limit, $N\to \infty,p\to\infty, p/N=\rho$,  we assume
that the roots of the Bethe equations (\ref{Bethe equations_gamma}) are
distributed in the complex plain along some continuous contour $\Gamma$ with
the analytical density $R(x)$, so that the sum of values of a function $%
\mathrm{f}(x)$ at the roots is given by 
\begin{equation}
\sum_{i=1}^{p}\mathrm{f}(x_i)=N\int_\Gamma \mathrm{f}(x)R(x)dx.
\end{equation}
After taking the logarithm and replacing the sum by the integral, the system
of equation (\ref{Bethe equations_gamma}) can be reformulated in terms of
single integral equation for the density. The particular solution
corresponding to the largest eigenvalue is specified by the appropriate
choice of branch of the logarithm. Then the integral equation should be
solved for a particular form of the contour, which is not known \textit{a
priori}, and being first assumed should be self-consistently checked after
the solution has been obtained. In practice, however, analytical solution is
possible in the very limited number of cases. Particularly, one,
corresponding to the contour closed around zero, yields the density 
\begin{equation}
R^{\left( 0\right) }(x)=\frac{1}{2\pi ix}\left( \rho -\sum_{n=1}^{\infty }%
\frac{x^{n}}{1-q^{n}}\right)  \label{R_0(x),|q|<1}
\end{equation}%
for $|q|<1$ and 
\begin{equation}
R^{\left( 0\right) }(x)=\frac{1}{2\pi ix}\left( \rho +\sum_{n=1}^{\infty }%
\frac{x^{n}}{1-q^{-n}}\right)  \label{R_0(x),|q|>1}
\end{equation}%
for $|q|>1$. This case corresponds to $\gamma =0$ and hence $\lambda (\gamma
)=0$. Since $R^{\left( 0\right) }(x)$ is analytic in the ring $0<|x|<1$, the
integration of it along any contour closed in this ring does not depend on
its form. Therefore, to fix the form of the contour additional constraints
are necessary. Such constraint appears if we require that the density
preserves its analyticity when $\gamma $ deviates from zero and the contour
becomes discontinuous. This constraint implies that the density $R^{\left(
0\right) }(x)$ vanishes at the break point $x_{c}$, which is a crossing
point of the contour $\Gamma$ and the real axis. 
\begin{equation}
R^{\left( 0\right) }(x_{c})=0  \label{x_c}
\end{equation}%
This equation was first obtained by Bukman and Shore as a conical point
condition for the asymmetric six-vertex model \cite{bs}. It is remarkable,
that after the resummation of series in (\ref{R_0(x),|q|<1},\ref%
{R_0(x),|q|>1}),  the equation (\ref{x_c}) coincides with eqs.(\ref{saddle
point, |q|>1},\ref{saddle point, |q|<1}) up to the replacements 
\begin{eqnarray}
x_c=z_0(1-q) \,\,\,\,\,\, \mathrm{and} \,\,\,\,\,\,\ x_c=z_0(q^{-1}-1)
\label{x_c-z_0}
\end{eqnarray}
for $|q|<1$ and $|q|>1$ respectively. Remember that $z_0$ was shown to
coincide with the speed $v$. It will be clear below that the same relation
between $v=(\lambda^{\prime}_0|_{\gamma=0})/N$ and $x_c$ follows directly
from the expression for $\lambda_0(\gamma)$ obtained from the Bethe
equations, without appealing to the partition function.

The further analysis is related with the calculation of finite size
corrections to the above expression of $R^{\left( 0\right) }(x)$, which
makes possible to probe into the nonzero values of $\gamma $. This can be
done with the help of method developed in \cite{kim,dl}. Its essential part,
is the construction of the inverse expansion $Z^{-1}(x)$ to the function of
the number of roots $Z(x)=\int^{x}R(x)dx$ near the break point $x_{c}$.
Since the derivative of $Z(x)$ in the thermodynamic limit vanishes (see eq.(%
\ref{x_c})), its inverse expansion reveals the square root singularity,
which in its turn becomes the origin of $1/\sqrt{N}$ terms in the finite
size expansion of $R(x)$. As a result we obtain the following parametric
dependence of \ $R\left( x\right) $ on $\gamma $, both being represented as
functions of the same parameter $C$. 
\begin{eqnarray}
R_{s} &=&R_{s}^{\left( 0\right) }-\frac{1}{N^{3/2}}\frac{1}{2\pi i}\frac{%
q^{|s|}}{1-q^{|s|}}  \notag \\
&\times &\sum_{n=0}^{\infty }\left( \frac{i}{2N}\right) ^{n}\frac{\Gamma (n+%
\frac{3}{2})}{\pi ^{n+\frac{3}{2}}}\frac{c_{2n+1,s}}{\sqrt{2i}}\mathrm{Li}%
_{n+\frac{3}{2}}(C) \\
\gamma  &=&\frac{1}{N^{3/2}}  \notag \\
&\times &\sum_{n=0}^{\infty }\left( \frac{i}{2N}\right) ^{n}\frac{\Gamma (n+%
\frac{3}{2})}{\pi ^{n+\frac{3}{2}}}\frac{\overline{c}_{2n+1}}{\sqrt{2i}}%
\mathrm{Li}_{n+\frac{3}{2}}(C),
\end{eqnarray}%
Here $R_{s}$ and $R_{s}^{\left( 0\right) }$ are the Loraunt coefficients of $%
R(x)$ and $R^{\left( 0\right) }(x)$ respectively defined as follows 
\begin{equation}
R(x)=\sum_{s=-\infty }^{\infty }R_{s}/x^{s+1},
\end{equation}%
$c_{2n+1,s}$ and $\overline{c}_{2n+1}$ are the coefficients of $x^{n}$ in $%
\left( \sum_{k=0}^{\infty }a_{k}x^{k}\right) ^{s}$ and $\log \left(
\sum_{k=0}^{\infty }a_{k}x^{k}\right) $ respectively, where $a_{n}$ are the
coefficients of the inverse expansion $Z^{-1}(x)$ near the point $x_{c}$.
The location of $x_{c}$ is to be self-consistently defined from the equation 
$R\left( x_{c}\right) =0$. For the first three orders of $1/\sqrt{N}$
expansion the\ coefficients $a_{n}$ can be obtained from the inverse
expansion of zero order function $Z^{\left( 0\right) }(x)=\int^{x}R^{\left(
0\right) }(x)dx$, while $R^{\left( 0\right) }(x)$ has been obtained above.
To evaluate the sum over the roots, one needs to integrate along the contour 
$\Gamma $, which can be obtained from the initial closed contour by cutting
out small segment connecting two roots closest to the point $x_{c}$. For a
function which is defined by the expansion 
\begin{equation}
\mathrm{f}(x)=\sum_{s=1}^{\infty }\mathrm{f}_{s}x^{s}
\end{equation}%
this yields 
\begin{equation}
\sum_{i=1}^{p}\mathrm{f}(x_{i})=2\pi iN\sum_{s=1}^{\infty }q^{\pm s}\mathrm{f%
}_{s}R_{s}
\end{equation}%
minus and plus in power of $q$ being for $|q|<1$ and $|q|>1$ respectively.
Finally the point $x_{c}$ enters to all the results trough the coefficients $%
c_{2n+1,s}$ and $\overline{c}_{2n+1}$. It is related with the physical
quantities through eqs.(\ref{saddle point, |q|<1},\ref{saddle point, |q|>1},%
\ref{v},\ref{x_c-z_0}). We use this relation, to write the final results as
a function of speed $v$, density $\rho $, and the parameter $q$. Below we
give the expression for the largest eigenvalue in the scaling limit $\gamma
N^{3/2}=const,N\rightarrow \infty $ 
\begin{equation}
\lambda _{0}(\gamma )=Nv\gamma +k_{1}G(k_{2}\gamma ).  \label{genfun}
\end{equation}%
Here the function $G(x)$ has the following parametric form 
\begin{eqnarray}
G(x) &=&-\mathrm{Li}_{5/2}(C) \\
x &=&-\mathrm{Li}_{3/2}(C)  \label{x(C)}
\end{eqnarray}%
with $\text{Li}_{\alpha }(x)=\sum_{k=1}^{\infty }x^{k}/k^{\alpha }$ is the
function of polylogarithm and the constants $k_{1},k_{2}$ are 
\begin{eqnarray}
k_{1} &=&\frac{1}{N^{3/2}}\sqrt{\frac{v/(1-q)}{8\pi }}\frac{g_{q}^{\prime
\prime }(v(1-q))}{(g_{q}^{\prime }(v(1-q)))^{5/2}} \\
k_{2} &=&N^{3/2}\sqrt{2\pi v(1-q)g_{q}^{\prime }(v(1-q))}
\end{eqnarray}%
for $|q|<1$, and 
\begin{eqnarray}
k_{1} &=&\frac{1}{N^{3/2}}\sqrt{\frac{v/(1-q^{-1})}{8\pi }}\frac{%
g_{1/q}^{\prime \prime }\left( v\left( q^{-1}-1\right) \right) }{\left(
g_{1/q}^{\prime }\left( v\left( q^{-1}-1\right) \right) \right) ^{5/2}} \\
k_{2} &=&-N^{3/2}\sqrt{2\pi v(1-q^{-1})g_{1/q}^{\prime }(v(q^{-1}-1))}%
\,\,\,\,\,\,\,\,\,\,\,\,\,\,\,
\end{eqnarray}%
for $|q|>1$, were 
\begin{equation}
g_{\nu }(x)=\sum_{n=1}^{\infty }\frac{x^{n}}{1-\nu ^{n}}.
\end{equation}%
The scaling form of the function $G(x)$ was suggested to be universal for
Kardar-Parisi-Zhang (KPZ) universality class \cite{kpz,da}. Using the
generating function obtained one can evaluate all the cumulants of the
travelled distance 
\begin{equation}
\lim_{t\rightarrow \infty }\frac{\langle Y_{t}^{n}\rangle _{c}}{t}=\left. 
\frac{\partial ^{n}\lambda _{0}(\gamma )}{\partial \gamma ^{n}}\right\vert
_{\gamma =0}.
\end{equation}%
The large deviation function, $\mathrm{ldf}(x)=\lim_{t\rightarrow \infty }{%
\log P(Y_{t}=x)}/{t},$ can be also obtained as a Legendre transformation of $%
\lambda _{0}(\gamma )$.

\section{Summary and discussion}

To summarize, we apply the Bethe ansatz to solve zero range process with the
totally asymmetric dynamics on a ring. The eigenfunctions of the master
equation have the form of the Bethe function weighted with the stationary
weights of corresponding particle configurations. The requirement of Bethe
ansatz integrability leads to the special choice of the rates of hopping of
particle out of a site. The rates should be $q$-numbers, $[n]_q$,
generalizing the case of noninteracting diffusing particles, where the rate
is equal to $n$, the number of particles at a site of departure. The
noninteracting case can be restored in the limit $q\to 1$. Two other
limiting cases, $q=0$ and $q\to \infty$ reproduce well known totally
asymmetric exclusion process and drop-push model respectively. The case of
general $q$ is shown to be equivalent to the q-boson totally asymmetric
diffusion model. Continuing analogy with noninteracting case, we show that
many quantities characterizing the stationary state correlations of the
model turns out $q$-analogs of corresponding functions appearing in the
noninteracting case. To provide an example of application of the Bethe
ansatz solution obtained, we derive the expression for the large time limit
of the generating function of cumulants of the total distance travelled by
particles. It has a universal form specific for KPZ universality class. The
question whether the q-boson totally asymmetric diffusion model belongs to
KPZ class was addressed in concluding remarks in \cite{sw2}. The result (\ref%
{genfun}-\ref{x(C)}) is an argument in favour of this assumption.

In connection with above results the following questions appear. First, is
it possible to generalize the proposed combination of the Bethe ansatz with
stationary weights to any other processes with nonuniform stationary state,
say asymmetric exclusion process with parallel update? The consideration of
the associated vertex models is also of interest. The different weights of
vertex configurations depending on the order of vertices would result in the
appearance of non-local interaction. Second, can one apply the matrix
product method to study the exclusion process with long range interaction
associated with zero range process considered here to probe into spatially
inhomogeneous situation, e.g. at the open chain. Appearance of $q$-numbers
seems to be an indication of this possibility. We expect that the matrix
product ansatz should be again appropriately weighted with stationary
weights of some homogeneous system. The consideration of such a system is
attractive, as the extra parameter $q$ could result in a reacher phase
diagram comparing to the usual totally asymmetric exclusion process. Third,
it is interesting to establish correspondence of the large scale behaviour
of the proposed process with KPZ equation.

\acknowledgments The author is grateful to V.B. Priezzhev and V.P.
Spiridonov for stimulating discussion. The work was supported partially by
FCT grant SFR/BPD/11636/2002 and by the grant of Russian Foundation for
Basic Research No.03-01-00780.

\end{document}